\documentstyle[11pt]{article}
\catcode`\@=11
\long\def\@makefntext#1{
\protect\noindent \hbox to 3.2pt {\hskip-.9pt
$^{{\ninerm\@thefnmark}}$\hfil}#1\hfill}		%CAN BE USED

\def\@makefnmark{\hbox to 0pt{$^{\@thefnmark}$\hss}}  %ORIGINAL

\def\ps@myheadings{\let\@mkboth\@gobbletwo
\def\@oddhead{\hbox{}
\rightmark\hfil\ninerm\thepage}
\def\@oddfoot{}\def\@evenhead{\ninerm\thepage\hfil
\leftmark\hbox{}}\def\@evenfoot{}
\def\sectionmark##1{}\def\subsectionmark##1{}}

\renewcommand{\thefootnote}{\fnsymbol{footnote}}

\newcounter{sectionc}\newcounter{subsectionc}\newcounter{subsubsectionc}
\renewcommand{\section}[1] {\vspace*{0.6cm}\addtocounter{sectionc}{1}
\setcounter{subsectionc}{0}\setcounter{subsubsectionc}{0}\noindent
	{\normalsize\bf\thesectionc. #1}\par\vspace*{0.4cm}}
\renewcommand{\subsection}[1] {\vspace*{0.6cm}\addtocounter{subsectionc}{1}
	\setcounter{subsubsectionc}{0}\noindent
	{\normalsize\it\thesectionc.\thesubsectionc. #1}\par\vspace*{0.4cm}}
\renewcommand{\subsubsection}[1]
{\vspace*{0.6cm}\addtocounter{subsubsectionc}{1}
	\noindent {\normalsize\rm\thesectionc.\thesubsectionc.\thesubsubsectionc.
	#1}\par\vspace*{0.4cm}}

\def\abstracts#1{{

\centering{\begin{minipage}{12.2truecm}\footnotesize\baselineskip=12pt\noindent
	\centerline{\footnotesize ABSTRACT}\vspace*{0.3cm}
	\parindent=0pt #1
	\end{minipage}}\par}}

\renewenvironment{thebibliography}[1]
	{\begin{list}{\arabic{enumi}.}
	{\usecounter{enumi}\setlength{\parsep}{0pt}
\setlength{\leftmargin 1.25cm}{\rightmargin 0pt}
	 \setlength{\itemsep}{0pt} \settowidth
	{\labelwidth}{#1.}\sloppy}}{\end{list}}

\newcounter{itemlistc}
\newcounter{romanlistc}
\newcounter{alphlistc}
\newcounter{arabiclistc}

 1
 1
 1

\font\ninerm=cmr9

\newcommand{\eq}[1]{(\ref{#1})}
\newcommand{\beq}{\begin{equation}}
\newcommand{\eeq}{\end{equation}}
\newcommand{\beqn}{\begin{eqnarray}}
\newcommand{\eeqn}{\end{eqnarray}}
\newcommand{\dd}{\mbox{d}}
\newcommand{\nsum}[2]{\sum_{\scriptstyle #1(\CK{#2})
\in \Z}}
\newcommand{\nddsum}[2]{\sum_{\stackrel{\scriptstyle \dual #1(\dual\CK{#2})
\in \Z} {\delta \dual #1=0}}}
\newcommand{\nDsum}[2]{\sum_{\scriptstyle \dual #1(\dual\CK{#2})
\in \Z}}
\newcommand{\nDDsum}[2]{\sum_{\stackrel{\scriptstyle #1(\CK{#2})
\in \Z} {\delta #1=0}}}
\newcommand{\A}{\theta}
\newcommand{\cR}{{\cal{R}}}
\newcommand{\cZ}{{\cal{Z}}}
\newcommand{\dual}{\mbox{}^{\ast}}
\newcommand{\intpi}{\int\limits_{-\pi}^{+\pi}}
\newcommand{\intinf}{\int\limits_{-\infty}^{+\infty}}
\newcommand{\Z}{{Z \!\!\! Z}}
\newcommand{\CK}[1]{\mbox{\scriptsize c}_{\mbox{$\scriptstyle #1$}}}
\newcommand{\dD}{{\cal D}}
\newcommand{\const}{{\rm const.} \cdot}

\def\bbbone{{\mathchoice {\rm 1\mskip-4mu l} {\rm 1\mskip-4mu l}
{\rm 1\mskip-4.5mu l} {\rm 1\mskip-5mu l}}}
\def\bbbc{{\mathchoice {\setbox0=\hbox{$\displaystyle\rm C$}\hbox{\hbox
to0pt{\kern0.4\wd0\vrule height0.9\ht0\hss}\box0}}
{\setbox0=\hbox{$\textstyle\rm C$}\hbox{\hbox
to0pt{\kern0.4\wd0\vrule height0.9\ht0\hss}\box0}}
{\setbox0=\hbox{$\scriptstyle\rm C$}\hbox{\hbox
to0pt{\kern0.4\wd0\vrule height0.9\ht0\hss}\box0}}
{\setbox0=\hbox{$\scriptscriptstyle\rm C$}\hbox{\hbox
to0pt{\kern0.4\wd0\vrule height0.9\ht0\hss}\box0}}}}
\def\bbbe{{\mathchoice {\setbox0=\hbox{\smalletextfont e}\hbox{\raise
0.1\ht0\hbox to0pt{\kern0.4\wd0\vrule width0.3pt
height0.7\ht0\hss}\box0}}
{\setbox0=\hbox{\smalletextfont e}\hbox{\raise
0.1\ht0\hbox to0pt{\kern0.4\wd0\vrule width0.3pt
height0.7\ht0\hss}\box0}}
{\setbox0=\hbox{\smallescriptfont e}\hbox{\raise
0.1\ht0\hbox to0pt{\kern0.5\wd0\vrule width0.2pt
height0.7\ht0\hss}\box0}}
{\setbox0=\hbox{\smallescriptscriptfont e}\hbox{\raise
0.1\ht0\hbox to0pt{\kern0.4\wd0\vrule width0.2pt
height0.7\ht0\hss}\box0}}}}

\def\NP{ Nucl.~Phys.}

\def\PL{ Phys.~Lett.}
\def\PRL{ Phys.~Rev.~Lett.}
\def\PRp{ Phys.~Rep.}
\def\PR{ Phys.~Rev.}

\def\dd{{\rm d}}

\def\cf{{\it cf.}}

\def\eg{{\it e.g.}}

\date{}
\begin{document}
\rightline{\normalsize ITEP--TH--12/95}
\rightline{\normalsize hep--lat/9512008}
\vspace{1cm}
\centerline{\normalsize\bf CONFINEMENT MECHANISM}
\baselineskip=22pt
\centerline{\normalsize\bf IN VARIOUS ABELIAN PROJECTIONS}
\centerline{\normalsize\bf OF LATTICE GLUODYNAMICS\footnote{
Talk given at the `Non-perturbative approaches to QCD'
workshop at ECT* in Trento.}
}
%\vfill
\vspace*{0.6cm}
\centerline{\footnotesize M.N.~CHERNODUB}
\baselineskip=13pt
\vspace*{0.3cm}
\centerline{\footnotesize\it ITEP, B.Cheremushkinskaya 25, Moscow,
117259, Russia}
\centerline{\footnotesize\it and}
\centerline{\footnotesize\it Moscow Institute of Physics and Technology,
Dolgoprudny, Moscow region, Russia}
\vspace*{0.3cm}
\centerline{\footnotesize M.I.~POLIKARPOV and A.I.~VESELOV}
\baselineskip=13pt
\vspace*{0.3cm}
\centerline{\footnotesize\it ITEP, B.Cheremushkinskaya 25, Moscow,
117259, Russia}
\vspace*{0.6cm}

\abstracts{We show that the monopole confinement mechanism in lattice
gluodynamics may be a particular feature of the maximal abelian projection.
We give an explicit example of the $SU(2) \rightarrow U(1)$ projection (the
minimal abelian projection), in which the confinement is due to topological
objects other than monopoles. We also discuss the string representation of
the abelian projected $SU(2)$ gluodynamics.}

\vspace*{0.6cm}
\normalsize\baselineskip=15pt
\setcounter{footnote}{0}
\renewcommand{\thefootnote}{\alph{footnote}}

\section{Introduction}
In his well known--paper, 't Hooft \cite{tHo81} suggested a
partial gauge fixing procedure for the $SU(N)$ gluodynamics which
does not fix the $[ U(1)]^{N-1}$ gauge group. Under the abelian
transformations, the diagonal elements of the gluon field transform
as gauge fields; the nondiagonal elements transforms as matter fields. Due
to the compactness of the $U(1)$ gauge group, the monopoles exist, and if
they are condensed, the confinement of color can be explained in the
framework of the classical equations of motion \cite{Man76,tHo76}. The
string between the colored charges is formed as the dual analogue of the
Abrikosov string in a superconductor, the monopoles playing the role of
the Cooper pairs.

        Many numerical experiments (see \eg\ the review \cite{Suz93})
confirm the monopole confinement mechanism in the $U(1)$ theory obtained by
the abelian projection from the $SU(2)$ lattice gluodynamics.  The string
tension $\sigma_{U(1)}$ calculated from the $U(1)$ Wilson loops (loops
constructed only from the abelian gauge fields) coincides with the full
$SU(2)$ string tension \cite{SuYo90}; the monopole currents satisfy the
London equation for a superconductor \cite{SiBrHa93}. Recently it has been
shown \cite{ShSu94,StNeWe94} that the $SU(2)$ string tension is well
reproduced by the contribution of the abelian monopole currents.  Numerical
study of the effective monopole action \cite{ShSu294} shows that the entropy
of the monopole loops dominates over the energy, and therefore, there exists
the monopole condensate in the zero temperature $SU(2)$ lattice
gluodynamics. All these remarkable facts, however, have been obtained only
for the so called maximal abelian (MaA) projection
\cite{KrLaScWi87}. Other abelian projections (such as the
diagonalization of the plaquette matrix $U_{x,12}$) do not give evidence
that the vacuum behaves as the dual superconductor. Below we give two
relevant examples.

        First, it turns out \cite{IvPoPo90} that the fractal dimensionality
of the monopole currents extracted from the lattice vacuum by means of the
maximal abelian projection is strongly correlated with the string tension.
If monopoles are extracted by means of other projections, this correlation is
absent (\cf\ Fig.2 and Fig.4 of ref.\cite{IvPoPo90}). Another example is
the temperature dependence of the monopole condensate measured on the basis
of the percolation properties of the clusters of monopole currents
\cite{IvPoPo93}. For the maximal abelian projection the condensate is
nonzero below the critical temperature $T_c$ and vanishes above it. For the
projection which corresponds to the diagonalization of $U_{x,12}$, the
condensate is nonzero at $T>T_c$, and it is not the order parameter for the
phase transition. The last result has been obtained by the authors of
\cite{IvPoPo93}, but is unpublished.

        In the present talk we discuss the dependence of the
confinement mechanism on the type of the abelian projection.  We find that
the monopole confinement mechanism is natural for the
MaA projection (Section~2), and we give an explicit example of the abelian
projection \cite{PolChe94} in which confinement is due to topological
defects which are not monopoles (Section~3).

\section{Maximal Abelian Projection and Compact Electrodynamics}

The MaA projection \cite{KrLaScWi87} corresponds to the gauge
transformation that makes the link matrices diagonal ``as much as
possible''.  For the $SU(2)$ lattice gauge theory, the matrices of the gauge
transformation $\Omega_x$ are defined by the following the maximization
condition:

%\samepage{
\beq
	\max_{\{\Omega_x\}} R(U')\;, \label{MaAP}
\eeq

\beq
        R(U') = \sum_{x,\mu} Tr(U'_{x\mu}\sigma_3 U'^{+}_{x\mu}\sigma_3)\;,
      \  U'_{x\mu} = \Omega^+_x U_{x\mu} \Omega_{x+\hat{\mu}}\;. \label{RU}
\eeq

For the standard parametrization of the $SU(2)$ link matrix, we have
$U^{11}_{x\mu} = \cos \phi_{x\mu} e^{i\theta_{x\mu}};$ $U^{12}_{x\mu} =
\sin \phi_{x\mu} e^{i\chi_{x\mu}};$ $ U^{22}_{x\mu} = U^{11 *}_{x\mu};$
$U^{21}_{x\mu} = - U^{12 *}_{x\mu};$ $ 0 \le \phi \le \pi/2,$ $ -\pi <
\theta,\chi \le \pi$; condition \eq{MaAP} has the form:

\beq
        \max_{\{\Omega_x\}}\sum_{x,\mu} \cos 2 \phi'_{x \mu}\;. \label{Mac}
\eeq
The $U(1)$ gauge transformations, which leave invariant the gauge conditions
\eq{MaAP}, \eq{Mac}, show that after the abelian projection $\theta$ becomes
the abelian gauge field and $\chi$ is the vector goldstone field, which
carry charge two in the continuum limit:

\beqn
        \theta_{x\mu} & \to & \theta_{x\mu} +\alpha_x -\alpha_{x+\hat{\mu}}
         \label{u1th}\;,\\
        \chi_{x\mu} & \to & \chi_{x\mu} +\alpha_x + \alpha_{x+\hat{\mu}}\;.
	  \label{u1chi}
\eeqn

It is instructive to consider the plaquette action in terms of the angles
$\phi, \ \theta $ and $\chi$:

\beq
S_P  =  \frac{1}{2}\mbox{Tr}\, U_1 U_2 U_3^+ U_4^+ = S^a + S^n + S^i\;,
\label{SP}
\eeq
where

\beqn
S^a  = & & \cos \theta_P\,
\cos\phi_1\, \cos\phi_2\, \cos\phi_3\, \cos\phi_4,
\nonumber \\
S^n  = & - & \cos (\theta_3 + \theta_4 - \chi_1 + \chi_2)\,
\cos\phi_3\, \cos\phi_4\, \sin\phi_1\, \sin\phi_2
\nonumber \\
 & + & \cos (\theta_2 + \theta_4 - \chi_1 + \chi_3)\,
                           \cos\phi_2\, \cos\phi_4\, \sin\phi_1\, \sin\phi_3
\nonumber \\
 & + & \cos (\theta_1 - \theta_4 + \chi_2 - \chi_3)\,
                           \cos\phi_1\, \cos\phi_4\, \sin\phi_2\, \sin\phi_3
\label{Sn} \\
 & + & \cos (\theta_2 - \theta_3 - \chi_1 + \chi_4)\,
                           \cos\phi_2\, \cos\phi_3\, \sin\phi_1\, \sin\phi_4
\nonumber \\
 & + & \cos (\theta_1 + \theta_3 + \chi_2 - \chi_4)\,
                           \cos\phi_1\, \cos\phi_3\, \sin\phi_2\, \sin\phi_4
\nonumber \\
 & - & \cos (\theta_1 + \theta_2 + \chi_3 - \chi_4)\,
                          \cos\phi_1\, \cos\phi_2\, \sin\phi_3\, \sin\phi_4,
\nonumber \\
 S^i  = & & \cos \chi_{\tilde{P}} \,
   \sin\phi_1\, \sin\phi_2\, \sin\phi_3\, \sin\phi_4; \nonumber
\eeqn
here we have set:

\beqn
\theta_P & =& \theta_1 + \theta_2 - \theta_3 - \theta_4\;, \label{P} \\
\chi_{\tilde{P}} & = & \chi_1 - \chi_2 + \chi_3 - \chi_4\;, \label{tP}
\eeqn
and the subscripts $1,...,4$ correspond to the links of the plaquette:  $1
\rightarrow \{x,x+\hat{\mu}\},...,4 \rightarrow \{x,x+\hat{\nu}$\}.  Note
that $S^a$ is proportional to the Wilson plaquette action of compact
electrodynamics for the ``gauge'' field $\theta$; the corresponding
action $S^i$ for the ``matter'' field $\chi$ contains the unusual
combination $\chi_{\tilde{P}}$ \eq{tP}, which is invariant under the gauge
transformations \eq{u1chi}. Action $S^n$ describes the interaction of the
fields $\theta$ and $\chi$.

         Due to condition \eq{Mac}, in the MaA projection the angle $\phi$
fluctuates near zero, and we can expect that the largest contribution to the
total action \eq{SP} comes from $S^a$, and  that $S^a > S^n > S^i$. This
conjecture is confirmed by numerical calculations. We use the standard heat
bath method to simulate $SU(2)$ gluodynamics on the
$10^4$ lattice, we study 15 values of $\beta$, $0.1 \le \beta \le 3.5$; at
each value of $\beta$ we used 15 field configurations separated by 100 of
heat bath sweeps. To obtain the MaA projection, we performed 800 gauge
fixing sweeps through the lattice for each field configuration. It occurs
that $<S^a>$ is close to the total action, the maximal difference between
$<S_P>$ and $<S^i>$ is at $\beta \approx 2.2$, where $<S^a> \approx 0.82
<S_P>$; $S^i$ is unexpectably small:  $<S^i> \approx - 0.001 \pm
0.0004$ at $\beta = 2.2$, at other values of $\beta$ the absolute value of
$<S^i>$ is even smaller. It is clear that if we neglect the fluctuations of
the angle $\phi$, as well as the Faddeev-Popov determinant, the $SU(2)$
action in the maximal abelian gauge is well approximated by the $U(1)$
action:  $S_P\approx\cos \theta_P$, with the renormalized constant
$\bar{\beta}=\beta \cos^4 \phi$.

Since in
the compact electrodynamics the confinement is due to the monopole
condensation, it is not surprising that in numerical experiments the vacuum
of gluodynamics behaves in the MaA projection as the dual superconductor. Of
course, this is only an intuitive argument. The confinement in the $U(1)$
theory exists in the strong coupling region, in which the rotational
invariance is absent.  Therefore, in order to explain the confinement at
large values of $\beta$ in $SU(2)$ gluodynamics, we have to study in detail
some special features of the gauge fixing procedure (such as the
Faddeev-Popov--determinant, fluctuations of the angle $\phi$, etc.).

        The fact that $<S^a>$ is close to $<S_P>$ is very interesting; it
means that in the MaA projection there is a small parameter in the $SU(2)$
lattice gluodynamics, which is $\varepsilon = \frac{<S_P>-<S^a>}{<S_P>}$; at
all values of $\beta$, we have $\varepsilon \le 0.18$. The meaning of this
parameter is simple: it is the natural measure of closeness between the
diagonal matrices and the link matrices after the gauge projection.
Therefore the lattice $SU(2)$ gauge theory in MaA projection is very close
to an abelian theory. In Appendix we show how the abelian theory with an
arbitrary action can be represented as a string theory. The strings carry
the electric flux, and confine electric charges.

\section{$SU(2)$ Gluodynamics in the Minimal Abelian Projection}

The minimal abelian (MiA) projection \cite{PolChe94} is defined similarly
to the MaA projection \eq{MaAP} by

\beq
        \min_{\{\Omega_x\}} R(U')\;, \label{MiAP}
\eeq
where $R(U')$ is defined by \eq{RU}. In this projection the largest part of
the plaquette action \eq{SP} is $S^i$, and the term which is most important
for the dynamics is $\cos \chi_{\tilde{P}}$ (rather than $\cos \theta_P$ as
it is in the MaA projection). The fields in the MiA projection can be
transformed into the fields in the MaA projection by the following
gauge transformation:

\beq
\Omega (x) = - i \sigma_2 \cdot \frac{(-1)^{x_1+x_2+x_3+x_4}+1}{2} +
\bbbone \cdot \frac{(-1)^{x_1+x_2+x_3+x_4} - 1}{2}\; . \label{Sigma2}
\eeq
Thus $\Omega(x)$ is equal to the unity in the
``odd'' sites of the lattice, and to $- i \sigma_2$ in the ``even'' sites;
this gauge transformation becomes singular in the continuum limit.
The angles
$\phi$, $\theta$ and $\chi$, which parametrize the link matrix $U_l$,
transform under this gauge transformation in the following way. If
the link starts at an even point, $\left((-1)^{x_1+x_2+x_3+x_4}=1\right)$,
then $U_l \rightarrow (-i \sigma_2) U_l$ and

\beq
\phi \rightarrow \frac{\pi}{2} - \phi, \ \theta \rightarrow -\chi, \ \chi
\rightarrow (\pi - \theta) \bmod 2\pi . \label{angl1}
\eeq
If the link starts at an odd point, then $U_l \rightarrow U_l (-i
\sigma_2)^+ $ and

\beqn
\phi \rightarrow \frac{\pi}{2} - \phi, \ \theta \rightarrow (\pi + \chi)
\bmod 2\pi, \ \chi \rightarrow \theta . \label{angl2}
\eeqn
Since $Tr(U'_{x\mu}\sigma_3 U'^{+}_{x\mu}\sigma_3) = \cos 2 \phi '$, it
follows that under this gauge transformation $$Tr(U'_{x\mu}\sigma_3
U'^{+}_{x\mu}\sigma_3) \to - Tr(U'_{x\mu}\sigma_3
U'^{+}_{x\mu}\sigma_3)\,,$$ and the fields in the MaA projection are
transformed into fields in the MiA projection (and vice versa). Moreover,
the monopoles extracted from the field $\theta$ in the MaA projection turn,
in the MiA projection, into some topological defects constructed from the
``matter'' fields~$\chi$. We call these topological defects ``minopoles''.

Minopoles can be extracted from a given configuration of gauge fields
similarly to monopoles: from the angles $\chi$ we construct gauge
invariant plaquette variables $\chi_{\tilde{P}} = \tilde{\dd}\chi \bmod 2\pi
$, where $\tilde{\dd}\chi$ is defined by \eq{tP}. From these plaquette
variables we construct the variables attached to the elementary cubes
${}^*j = \frac{1}{2\pi}\tilde{\dd} \chi_{\tilde{P}}$; for ${}^*j \neq 0$
the link dual to the cube carries the minopole current. We use the
notation $\tilde{\dd}$ (instead of $\dd$), since the gauge transformations
of $\chi$ given by \eq{u1chi} differ from the gauge transformations of
$\theta$ given by \eq{u1th}, and the construction of the plaquette variable
from the link variables and that of the cube variable from the plaquette
variables differ in an obvious way from the standard construction. For
example, $\dd\theta$ is defined by \eq{P} and $\tilde{\dd}\chi$ is defined
by \eq{tP}. In Fig.1 we illustrate the standard construction of the
monopoles from the field $\theta$, and the construction of the minopoles
from the field $\chi$.

Since monopoles, which exist in the MaA projection become minopoles in the
MiA projection, than if in the MaA projection the confinement phenomenon is
due to condensation of monopoles (constructed from the field $\theta$), then
in the MiA projection the confinement is due to other topological objects
(minopoles), constructed from the ``matter'' field $\chi$. We thus conclude
that {\em in the MiA projection the confinement is not due to monopoles and
the vacuum is not an analogue of the dual superconductor}. It should be
stressed that monopoles still exist in the MiA projection; they can be
extracted from the fields $\theta$ in the usual way, but they are not at all
related to the dynamics. To illustrate this simple fact we plot in Fig.2 the
space--time asymmetry of the {\bf \it monopole} currents
\cite{Bra91Hio91}\footnote{The definition of this asymmetry
is obvious:
$A=<(J_t-J^S)/J_t>$, where $J^S= (J_x+J_y+J_z)/3$, $J_{\mu}$ is the monopole
current in the direction $\mu$.} \, for the $SU(2)$ gauge theory on the
$10^3\times4$ lattice for the MiA projection. In the same figure we also
show the asymmetry of the minopole currents in the MiA projection. It is
well known that the temperature phase transition is at $\beta\approx 2.3$
for the $10^3\times4$ lattice. It is clearly seen that the asymmetry of the
minopole currents is the order parameter for the temperature phase
transition, while the asymmetry of the monopole currents is not. Since the
monopole currents and the minopole currents are interchanged when the fields
are transformed from the MiA to the MaA projection, Fig.2 also shows that
for the MaA projection the asymmetry of the monopole currents is the order
parameter, whereas the asymmetry of minopole currents is not an order
parameter. These results have been obtained by averaging over 10
statistically independent field configurations for each value of $\beta$,
and 500--800 of gauge fixing sweeps have been performed for each
configuration.

        Minopoles are to some extend the lattice artifacts, since the gauge
fields in the MaA and the MiA projections are related by the gauge
transformation, which becomes singular in the continuum limit. We discuss
minopoles, since they clearly illustrate the dependence of the confinement
mechanism on the lattice, upon the type of the abelian projection.

\newpage
\section{Conclusions}

        If monopoles are responsible for the confinement in the MaA
projection and minopoles are responsible for the confinement in the MiA
projection, what are the important topological excitations in a general
abelian projection? If both diagonal and nondiagonal gluons are not
suppressed, then, in plus to monopoles and to minopoles, string--like
topological defects can also be important for the dynamics of the system
\cite{ChPoZu94}. The idea is: nondiagonal gluons transform under the $U(1)$
gauge transformations as matter fields, diagonal gluons transform as gauge
fields, and an analogue of the Abrikosov--Nielsen--Olesen strings exists in
gluodynamics after the abelian projection. Between strings made of condensed
nondiagonal gluons (which carry the $U(1)$ charge {\bf 2}) and the test
quark of the charge {\bf 1}, there exists topological interaction
\cite{AhBheff,PoWiZu93}, which is the analogue of the Aharonov -- Bohm
effect. Thus, in the effective $U(1)$ action of the $SU(2)$ gluodynamics
there probably exists a very specific topological interaction. We describe
an analytical and numerical study of this interaction in a separate
publication.

The topological defects discussed above may be a reflection of some $SU(2)$
gauge field configuration. For example, monopoles and minopoles may be the
abelian projection of $SU(2)$ monopoles \cite{SmvdSi93}. In ref.
\cite{TrPoWo93} it is found that the ``extended monopoles'' \cite{IvPoPo90}
may be important for the confinement mechanism in different abelian
projections of the $3D$ $SU(2)$ gluodynamics. Finally, we note that it was
found recently \cite{SuIlMaOkYo94} that the contribution of the Dirac sheets
to the abelian Polyakov loops plays the role of the order parameter for
finite temperature lattice gluodynamics; it is interesting that this fact
holds not only for the MaA projection but also for others unitary gauges.
It means that in the considered unitary gauges monopoles are important for
the dynamics, other topological excitations (minopoles and strings) may be
also important.  Note that in the MiA projection monopoles are substituted
by minopoles, and in this projection we expect that the contribution of the
minopole ``Dirac sheets'' is correlated with the expectation values of the
Polyakov loops.

\vspace*{0.6cm}
\noindent
{\normalsize\bf Acknowledgments}\par\vspace*{0.4cm}

Authors are grateful to E.~Akhmedov, P.~van~Baal, R.~Haymaker, Y.~Matsubara,
J.~Smit, T.~Suzuki
and K.~Yee for interesting discussions. This work was supported by the grant
number MJM000, financed by the International Science Foundation, by the JSPS
Program on Japan -- FSU scientists collaboration and by the grant number
93-02-03609, financed by the Russian Foundation for the Fundamental Sciences.

\newpage
\vspace*{0.6cm}
\noindent
{\normalsize\bf Appendix A}\par\vspace*{0.4cm}

First we perform the analogue of the Berezinski--Kosterlitz--Thauless (BKT)
transformation \cite{BKT} for an arbitrary $U(1)$ action $S[\dd \A]$,
$S[\dots, X_P+2 \pi, \dots] = S[\dots, X_P, \dots]$, where $P$ denotes any
plaquette of the original lattice. For the sake of convenience we use the
differential forms formalism on the lattice (see Ref.\cite{Intro} for the
introduction). We start from the partition function of the compact $U(1)$
theory

\beq
        \cZ = \intpi \dD \A \exp\Bigl\{- S[\dd \A] \Bigr\}\,.
                        \label{initial}
\eeq
The expansion of the function $e^{- S(X)}$ in
the Fourier series yields:

\beq
     \cZ = \const \intpi \dD \A \nsum{n}{2}
     F(n) e^{i (n,{\rm d} \A) }\,, \label{sum}
\eeq
where $n$ is an integer--valued two-form,

\beq
        F(n) = \const \intpi \dD X \exp\Bigl\{- S[X]
        -i (n, X) \Bigr\}\,. \label{FT}
\eeq
 Let us insert the unity

\beq
        1 = \intinf \dD G \, \delta(G-n) \nonumber
\eeq
        into the sum (\ref{sum}):
\beq
      \cZ = \intpi \dD A \exp\Bigl\{- S_P(\dd A) \Bigr\} = \const
      \intpi \dD A \intinf \dD G \nsum{n}{2} \delta(G-n)
      F(G) e^{i (G,\dd A )}
\eeq
and use the Poisson summation formula

\beq
       \nsum{n}{2} \delta(G-n) = \nsum{n}{2} e^{2 \pi i (G, n)}\,,
\eeq
the result is:

\beq
     \cZ = \const \intpi \dD \A  \intinf \dD G \nsum{n}{2}
     F(G)\, e^{i ({\rm d} \A+ 2 \pi n, G)}\,. \label{poisson}
\eeq
Here $G$ is a real--valued two--form.

Now we perform the BKT transformation with respect to the
integer--valued 2-form~$n$:

\beq
n = m[j] + \dd q\,,\quad \dd m[j] = j\,,\quad \dd j =0\,, \label{bkt0}
\eeq
where $q$ and $j$ are one-- and three--forms respectively. First change the
summation variable, $\displaystyle{\nsum{n}{2} =
\sum_{\stackrel{\scriptstyle j(\CK{3}) \in \Z} {\dd j=0}}
\nsum{q}{1}}$. Using the Hodge--de--Rahm decomposition we adsorb the
d--closed part of the 2--form $n$ into the compact variable $\A$:

\beq
  \dd \A + 2 \pi n = \dd \A_{n.c.} + 2 \pi \delta {\Delta}^{-1} j\,,
  \quad \A_{n.c.} = \A + 2 \pi
  {\Delta}^{-1} \delta m[j] + 2 \pi q\,. \label{bkt}
\eeq

Substituting eq.(\ref{bkt}) in eq.(\ref{poisson}) and integrating out the
noncompact field $\A_{n.c.}$ we get the following representation of the
partition function:

\beq
        \cZ = \const \intinf \dD G \nddsum{j}{1}
        F(G)\, \exp\Bigl\{2 \pi i (G, \delta
        {\Delta}^{-1} j) \}\, \delta(\delta G)\,. \label{step3}
\eeq

Let us consider the lattice with the trivial topology (e.g. $\cR^4$). Then
the constraint $\delta G = 0$ can be solved as\footnote{If we consider the
space with a nontrivial topology, arbitrary harmonical forms must be added
to the r.h.s. of this equation.}\,\, $G = \delta H$ where $H$ is a real
valued 3--form.  Substituting this solution into eq.(\ref{step3}) we obtain
on the dual lattice the final expression for the BKT--transformed action:

\beq
        \cZ = \const \nddsum{j}{1} e^{- S_{mon}(\dual j)}\,.
        \label{GenFin:1}
\eeq
where

\beq
        S_{mon}(\dual j) = - \ln \left(\,\, \intinf \dD H F(\delta H)
        \exp\Bigl\{2 \pi i (\dual H,\dual j) \Bigr\} \right)\,,
        \label{GenFin:2}
\eeq
we used the relation $\dd \delta {\Delta}^{-1} j \equiv j,\ \forall j : \dd
j = 0$.  Therefore for the general $U(1)$ action $S[\dd \theta]$ the
monopole action \eq{GenFin:2} is nonlocal and is expressed through the
integral over all lattice ($\intinf \dD H$). It is well known that for the
Villain form of the $U(1)$ action

\beqn
   \exp\{-S[{\rm d} \theta]\} = \nsum{n}{2}
   \exp\left\{ - \beta {||{\rm d} \theta + 2 \pi n||}^2 \right\}\,,
   \label{Villain}
\eeqn
the monopole action has the simple form: $S_{mon}(\dual j) = 4 \pi^2 \beta
(\dual j, \Delta^{-1} \dual j)$.

The partition function (\ref{GenFin:1},\ref{GenFin:2}) can be rewritten as:

\beq
        \cZ = \const \intinf \dD H \intpi \dD \vartheta
        F(\delta H) \nDsum{j}{1} \exp\left\{ i (\dd \dual \vartheta
        + 2 \pi \dual H,\dual j) \right\}\,,
        \label{close}
\eeq
where we introduced the compact fields $\vartheta$ to represent the
closeness of the monopole currents~$\dual j$. The use of the Poisson formula
leads to

\beq
        \cZ = \const \intinf \dD H \intinf \dD C \intpi \dD \vartheta
        F(\delta H) \nDsum{j}{1} \exp\left\{i (\dd \dual \vartheta +
        2 \pi \dual H + 2 \pi \dual j, \dual C) \right\}\,,
        \label{DefLeppard}
\eeq
where the $\delta$--functions are represented as the integral over the new
field $C$. Let us perform the BKT transformation with respect to the
integer--valued one--form $\dual j$. Repeating all the transformations
which leaded as to eq.\eq{bkt} we get

\beqn
      \dd \dual \vartheta + 2 \pi \dual j =
      \dd \dual \vartheta_{n.c.} + 2 \pi \delta \Delta^{-1} \dual \sigma\,,
      \qquad \delta \sigma = 0\,,
      \label{BKT2}
\eeqn
$\vartheta_{n.c.}$ is new noncompact field. Substituting eq.\eq{BKT2} in
eq.\eq{DefLeppard} and integrating over the field $\vartheta_{n.c.}$ we get
the constraint on the field $C$, $\delta \dual C \equiv \dual \dd C = 0$,
which can be resolved by the introduction of new noncompact field $C = \dd
B$. Using the identity $(\delta \Delta^{-1} \dual \sigma, \dual \dd B) =
(\sigma, B)$ (which is valid if $\delta \sigma = 0$) we get:

\beqn
     \cZ = \const \intinf \dD H \intinf \dD B \nDDsum{\sigma}{2}
     F(\delta H) \exp\left\{ 2 \pi i (H, \dd B)
     + 2 \pi (\sigma, B) \right\}\nonumber\\
     = \const \intinf \dD B \nDDsum{\sigma}{2}
     \exp\left\{ - S_{KR}(B) + 2 \pi i (\sigma, B) \right\} \nonumber\\
     = \const \nDDsum{\sigma}{2}
     \exp\left\{ - S_{str}(\sigma) \right\}
     \,,\label{Sound}
\eeqn
where $S_{KR}(B)$ is the action for the Kalb--Ramond lattice fields $B$,

\beqn
        S_{KR}(B) = - \ln \left(\,\, \intinf \dD H
        F(\delta H) \exp\left\{ 2 \pi i (H, \dd B)\right\} \right)\,,
        \label{Kalb}
\eeqn
the term $(\sigma, B)$ represents the interaction between the string
world sheet $\sigma$ and the Kalb--Ramond fields $B$.

The string action results from the integration over the field $B$:

\beqn
    S_{str}(\sigma) = - \ln \left(\,\, \intinf \dD B
    \exp\left\{ - S_{KR}(B) + 2 \pi i (\sigma, B) \right\} \right)\,.
\eeqn

It can be shown that the strings $\sigma$ carry electric fluxes. Therefore
the partition function of the compact electrodynamics is reduced to the sum
over the electric strings. The string action is complicated in general case,
but if we start with the Villain $U(1)$ \eq{Villain} action we get the
simple expression for the string partition function:

\beq
    \cZ = \const \nDDsum{\sigma}{2}
     \exp\left\{ - \frac{1}{4 \beta} {|| \sigma ||}^2 \right\}\,.
\eeq

\vspace*{0.6cm}
\noindent
{\normalsize\bf Figure Captions}\par\vspace*{0.4cm}
Fig.1. Construction of a monopole from the field $\theta$ and minopole from
the field $\chi$.

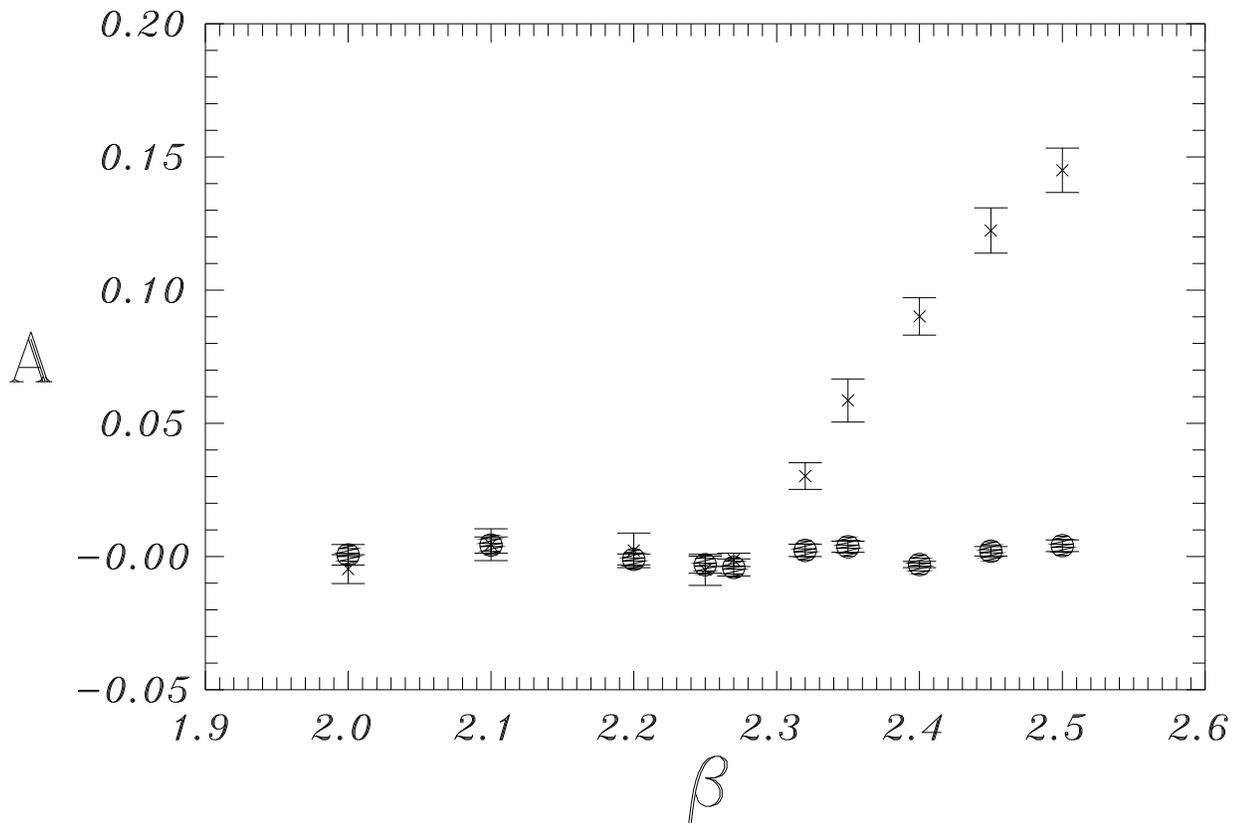
Fig.2. Asymmetry of the monopole currents (circles) and the minopole
currents (crosses) in the minimal abelian projection.

%\vspace*{0.4cm}

\newpage
\noindent
{\normalsize\bf References}\par

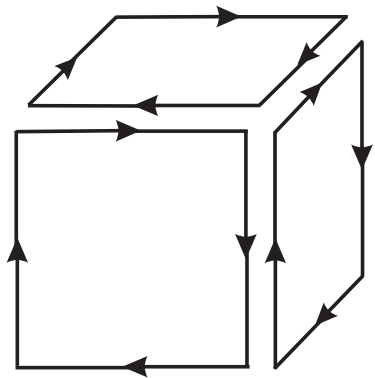
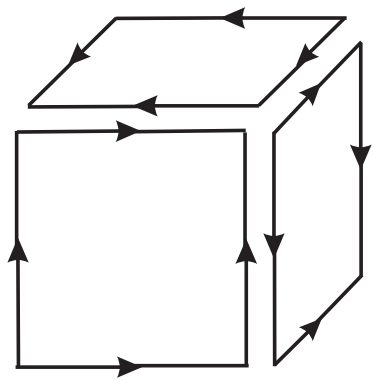
\end{document}